\documentclass[aps,pra,a4paper,showpacs,twocolumn]{revtex4}

\input epsf

\usepackage[cp1250]{inputenc} 
\usepackage[T1]{fontenc} 

\usepackage{amssymb}
\usepackage{amsmath}
\usepackage{amsfonts}
\usepackage{graphicx}
\usepackage{epsfig}
\usepackage{multirow}

\def\beq{\begin{equation}}
\def\eeq{\end{equation}}
\def\beqa{\begin{eqnarray}}
\def\eeqa{\end{eqnarray}}
\def\fig{Fig. }

\newcommand{\ket}[1]{\mbox{$ | #1 \rangle $}}
\newcommand{\bra}[1]{\mbox{$ \langle #1 | $}}
\newcommand{\eq}[1]{Eq.(#1) }

\begin{document}

\renewcommand{\figurename}{Figure}

\title{Quantum repeaters based on heralded qubit amplifiers}
\date{\today}

\author{Ji\v{r}í Miná\v{r}$^{1,2}$, Hugues de Riedmatten$^{3,4},$ and Nicolas Sangouard$^{2}$}
\affiliation{$^{1}$Centre for Quantum Technologies, National University of Singapore, 3 Science drive 2, Singapore 117543 \\
$^{2}$Group of Applied Physics, University of Geneva, CH-1211 Geneva 4, Switzerland\\
$^{3}$ICFO-Institute of Photonic Sciences, Mediterranean Technology Park, E-08860 Castelldefels (Barcelona), Spain \\
$^{4}$ICREA-Institució Catalana de Recerca i Estudis Avançats, E-08015 Barcelona, Spain}

\begin{abstract}
We present a quantum repeater scheme based on the recently proposed qubit amplifier [N. Gisin, S. Pironio and N. Sangouard, Phys. Rev. Lett. 105, 070501 (2010)]. It relies on a on-demand entangled-photon pair source which uses on-demand single-photon sources, linear optical elements and atomic ensembles. Interestingly, the imperfections affecting the states created from this source, caused e.g. by detectors with non-unit efficiencies, are systematically purified from an entanglement swapping operation based on a two-photon detection. This allows the distribution of entanglement over very long distances with a high fidelity, i.e. without vacuum components and multiphoton errors. Therefore, the resulting quantum repeater architecture does not necessitate final postselections and thus achieves high entanglement distribution rates. This also provides unique opportunities for device-independent quantum key distribution over long distances with linear optics and atomic ensembles. 
\end{abstract}

\maketitle
\section{Introduction}
\label{Introduction}

When tackling the task of distributing entanglement over long distances, the main problem comes from the transmission losses. In contrast to classical communications, the losses cannot merely be compensated by a straightforward amplification, since the cloning of an unknown single quantum state is impossible \cite{Wootters82}. Briegel et al. \cite{Briegel98} proposed to repeatedly profit from the fascinating feature of entanglement that it can be swapped \cite{Zukowski93}. The idea consists first to divide the total distance into smaller pieces (called elementary links), to distribute the entanglement in each of the elementary links, and to swap successively the entanglement between the adjacent links until it is extended over the whole distance. The resulting tool, the so-called quantum repeater, provides entanglement distribution rates that scale much better with the distance than the direct transmission of photons through optical fibers. However, the price to pay for its implementation is to be able to create the entanglement remotely in a heralded way, to store it and to swap it many times. \\

Duan et al. \cite{Duan01} have shown how to meet all the above described requirements using simple ingredients, i.e. atomic ensembles, linear optical elements and photon counting techniques. Although  this protocol has inspired a lot of experiments (see e.g. \cite{Sangouard11} for a recent review), it requires prohibitively long storage times to outperform the direct transmission of photons, even under optimistic assumptions on the memory and detection efficiencies. Many improvements have recently emerged \cite{improvementsQR, Chen07, Sangouard08}, which in turn spurred comparison between various architectures \cite{Sangouard11} and led to the conclusion that the local creation of high-fidelity entangled pairs of atomic excitations with the use of two-photon detections for long-distance entanglement generation permits the implementation of much more efficient quantum repeaters \cite{Sangouard08, Sangouard11} with the same ingredients.\\

Following these lines, we present an approach inspired by the heralded qubit amplifier recently proposed in Ref. \cite{Gisin10}, to produce entanglement on-demand using a probabilistic entangled-pair source, on-demand single-photon sources, linear optics and ensembles of atoms as quantum memories. Interestingly, we show that the imperfections resulting from the use of detectors and single-photon sources with non-unit efficiencies, can all be perfectly purified from an entanglement swapping based on a two-photon detection. Therefore, a state with a very high fidelity can be created in the elementary link and we show that the fidelity is preserved when the state is further extended over long distances using many entanglement swapping operations. In contrast to all previously proposed protocols based on linear optics and atomic ensembles, the final state does not contain vacuum components even in the presence of memory and detector imperfections, and hence, no post-selection is needed. This leads to high entanglement distribution rates and opens the way for applications where the postselections are not allowed, such as device-independent quantum key distribution (see for the principle \cite{diqkdprinciple} and for possible practical realizations \cite{Gisin10, diqkdrealization, Curty11, Pitkanen11} . \\

The paper is organized as follows. In Section \ref{sec QR local}, we first present an ideal quantum repeater architecture, i.e. involving a photon-pair source producing entanglement in an on-demand way and for which, the achievable entanglement distribution rate is only limited by the detection, storage and transmission inefficiencies. In the following sections, we show how such a quantum repeater can be approached using qubit amplifiers. The section \ref{subsec Local generation} shows how a source producing entanglement in an on-demand way can be realized by modifying the qubit amplifier scheme of Ref. \cite{Gisin10}. In \ref{subsec ent creation} and \ref{subsec Swapping}, we show how the created entanglement can be purified when it is processed to be shared between remote locations and how it can further be swapped between adjacent links to be extended over very long distances. Finally, in section \ref{subsec Results}, we illustrate numerically the performance of the proposed quantum repeater for a typical distance of 1000 km by taking the main experimental imperfections into account. The last section \ref{sec Conclusion} is devoted to the conclusion.\\

\section{Quantum repeater with on-demand entangled-photon pair sources}
\label{sec QR local}

In this section, we start by recalling the basic formula that is used to evaluate the achievable entanglement distribution rate with quantum repeaters. We first use this formula to quantify the performance of an ideal quantum repeater serving as a reference in the rest of the manuscript, which would use hypothetical sources producing maximally entangled states with unit efficiency. We reuse it in the last paragraph to show that it is essential to herald the creation of the entanglement in the elementary link.\\

The average time for the distribution of one entangled pair that can be achieved with a quantum repeater made with $n$ nesting levels ($2^n$ elementary links) is well approximated by (see appendix of Ref. \cite{Sangouard11})
\beq
	T_{tot} = \left( \frac{3}{2} \right)^n \frac{L_0}{c}\frac{1}{P_0 P_1 .. P_n},
	\label{eq Ttot_basic}
\eeq
where $L_0$ is the length of the elementary link, $P_0$ is the probability of a successful entanglement creation  in the elementary link and $P_k$ with $k=1,2,...$ is the success probability of the $k-$th entanglement swapping.\\

\begin{center}
	\begin{figure}[h!]
  	\includegraphics[width=6cm]{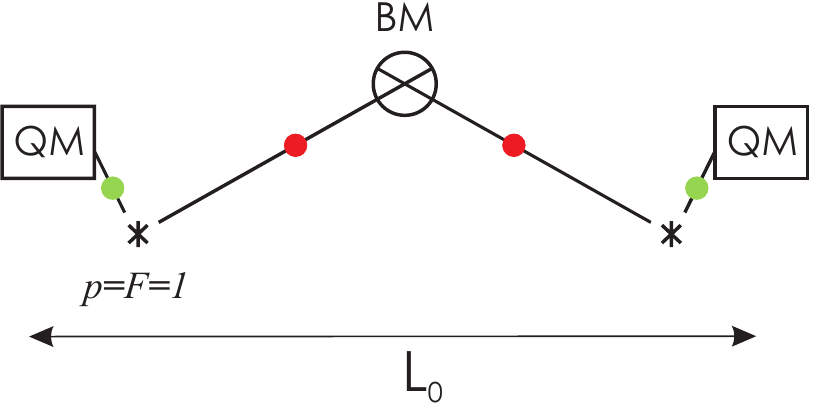} 
  	\caption{(Color online) Entanglement creation at a distance using hypothetical pair sources, producing on-demand maximally entangled states. Star - ideal source of entangled pairs, QM - quantum memories, BM - Bell measurement.} 
  	\label{fig1}
 	\end{figure}
\end{center}

It is instructive to calculate the entanglement distribution rate that could be achieved considering that on-demand photon-pair sources producing entanglement with unit fidelities are available. The basic procedure for entanglement creation between remote locations would require one source and one quantum memory at each location (c.f. Fig. \ref{fig1}). The two ideal sources would be excited simultaneously. One mode of the produced entanglement would be stored locally whereas the twin mode would be sent at a central station where a Bell measurement would be performed. The success of the Bell measurement would lead to the entanglement of the excitations stored into the memories. The entanglement could be extended by releasing the photons stored at the same location and then, performing successive entanglement swapping operations using linear optical elements and photon counters. Such a repeater would only be limited by detection and memory imperfections. Let $\eta_D$ and $\eta_M$ be the corresponding efficiencies. Label $\eta_t$ the transmission efficiency from one location to the central station where the Bell measurement is performed. ($\eta_t={\rm exp}(-L_0/(2L_{att})$ where $L_0$ is the length of one elementary link and $L_{att}$ is taken to be equal to $22$ km corresponding to a standard telecommunication fiber.) The probability of a successful entanglement creation in the elementary link would be given by $P_0=\frac{1}{2}(\eta_D\eta_t)^2.$ The factor 1/2 comes from the fact, that we are performing a photonic Bell measurement with linear optics. The square results from a Bell measurement based on a twofold coincidence detection, c.f. below. Similarly, the probability for the $k-$th swapping to succeed is $P_k = \frac{1}{2} (\eta_D \eta_M)^2$ for $k=1,2,...$. The average entanglement distribution time, in this ideal case, would thus be given by
\beq	
	T_{tot,id} = 2 \cdot 3^n \frac{L_0}{c}\frac{1}{\eta_D^{2n+2} \eta_M^{2n} \eta_t^2}.
	\label{eq Ttot_id}
\eeq
For 1000 km, $\eta_D=\eta_M=0.9$ and $n=4,$ this would translate into the distribution of one entangled pair every $6$ s. Quantum memories with comparable storage time may be within reach, given the recent experimental progresses in coherent optical memories \cite{storage_time} with demonstrated storage times in the second range (albeit in the classical regime).\\

\begin{center}
	\begin{figure}[h!]
  	\includegraphics[width=6cm]{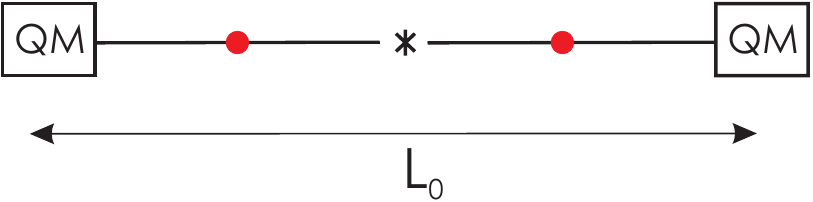} 
  	\caption{(Color online) Entanglement creation at a distance without heralding. Star - ideal source of entangled pairs, QM - quantum memory.} 
  	\label{fig2}
 	\end{figure}
\end{center}

Note that the process (the Bell measurement) heralding the successful creation of entanglement within each elementary links is crucial. To emphasize on this point, we give a short counter-example. Consider that each elementary link is made with two distant memories and one entanglement source localized half-way between the memories, c.f. Fig. \ref{fig2}. Once the source is excited, the photons propagate through lossy optical fibers and hopefully, reach the memories where they are stored. As before, two adjacent memories sitting at the same site are read out so that the entanglement can be swapped between the links. The swapping procedure is repeated until the entanglement reaches the desired distance. In opposition to the previously described protocol, there is now no announcement to herald the success of the entanglement creation in each link. Therefore, the distributed state has a vacuum component (assuming channel transmission typically small compared to one) that is large compared to the weight of the entangled component. Hence, a final purification is required to get rid of the vacuum. It is performed post-selectively by converting the memory excitations back into photons and by verifying that there is one photon in each location. Measurements in arbitrary basis are possible by choosing the appropriate setting. After a straightforward calculation, one finds that the average time to distribute entanglement with this protocol would be of the order of $10^5$ s \cite{note_fid}, roughly two orders of magnitude longer than for the original protocol by Duan, Lukin, Cirac and Zoller \cite{Duan01}.\\

These considerations remind us that efficient quantum repeaters necessitate the use of on-demand entangled-pair sources and the creation of remote entanglement in a heralded way. We describe in the following an attractive photon-pair source producing states with a high fidelity and where the residual errors can be purified by the operation heralding the creation of entanglement at a distance. This leads to a repeater architecture achieving unequaled entanglement distribution rates with atomic ensembles and linear optics.

\section{On-demand entanglement source based on a qubit amplifier}
\label{subsec Local generation}

\begin{center}
	\begin{figure}[h!]
  	\includegraphics[width=9cm]{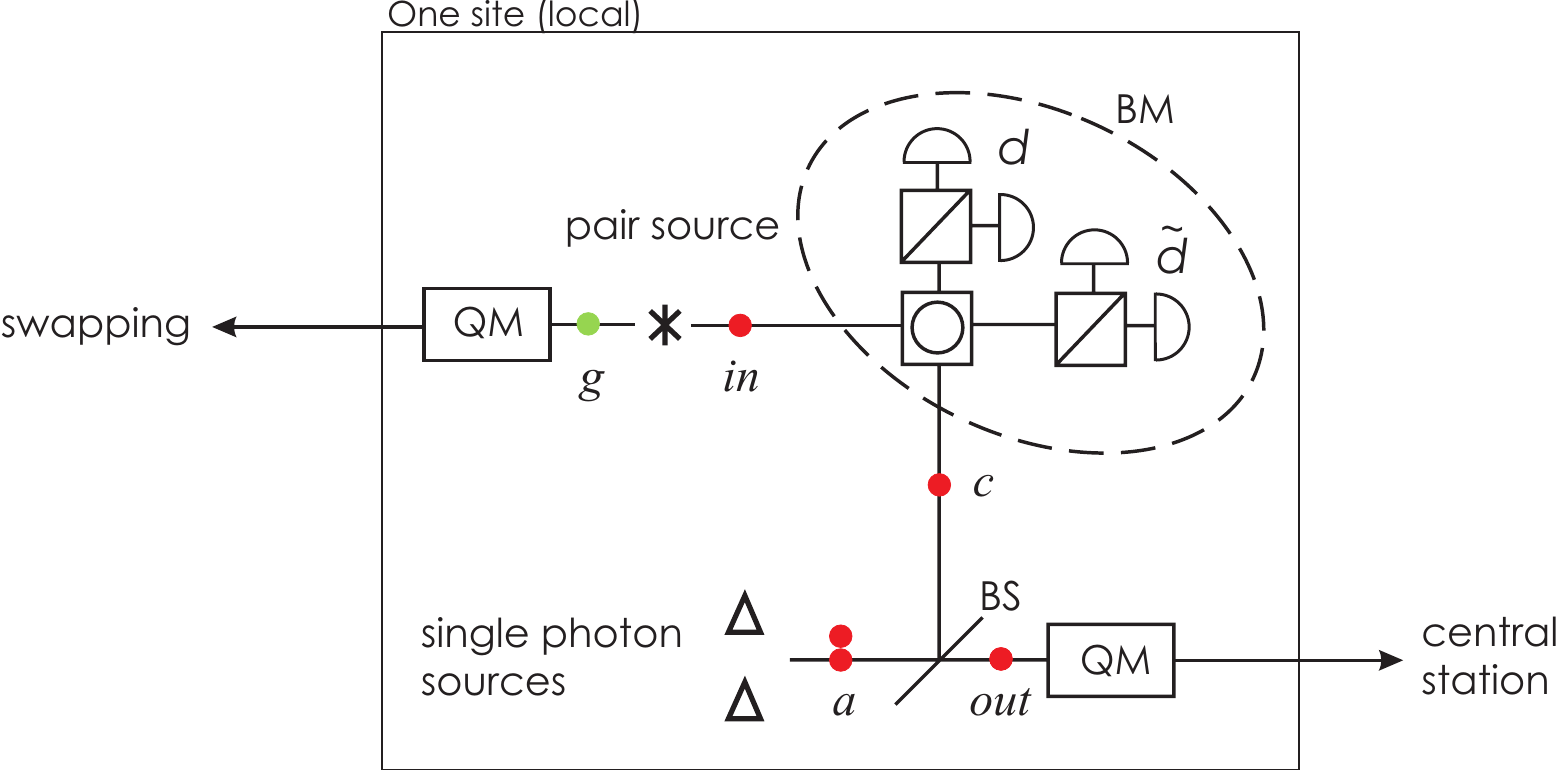} 
  	\caption{(Color online) Heralded source of entangled pairs, QM - quantum memories, BM - Bell measurement. A pair source (star) produces a polarization entangled state in a probabilistic way in modes $in$ and $g.$ Two single-photon sources (triangles) produce a product state of two photons with orthogonal polarizations in the same spatial mode $a$. The mode $a$ is sent through a the beam splitter (BS) with tunable reflectivity $R$ (in intensity) and the two photon coincidence detection $d$-$\tilde{d}$ teleports the mode $in$ to the mode $out$ (up to a unitary transformation). This leads to the entanglement between the modes $g$ and $out$ which are then stored in the memories. The BM here is inspired by \cite{Chen07} and consists of one PBS in the $\pm 45^o$ basis followed by a PBS in H/V basis at each output. } 
  	\label{fig QA}
 	\end{figure}
\end{center}

The principle of the proposed on-demand source is shown in \fig{\ref{fig QA}}. It starts with a source producing pairs of photons entangled in polarization, in a probabilistic way, meaning that each time it is excited, it emits an entangled pair with a small probability $p,$ corresponding to the state
\begin{eqnarray}
\rho_{pair} &=& (1-p)\ket{0}\bra{0} + \\
\nonumber
&& \frac{p}{2} (g^\dagger_{\rm{H}} in^\dagger_{\rm{H}} + g^\dagger_{\rm{V}} in^\dagger_{\rm{V}})\ket{0}\bra{0}(g_{\rm{H}} in_{\rm{H}} + g_{\rm{V}} in_{\rm{V}}).
\end{eqnarray}
$g$ and $in$ label two spatial modes. The polarization of these modes is given by the subscript. The state 
$g^\dagger_{\rm H} |0\rangle\langle 0| g_{\rm H}$ is, for example, associated to one excitation in mode $g$ with the horizontal polarization. Inspired by a recent proposal by Ralph and Lund \cite{Ralph09}, it has been shown how a suitable teleportation of the mode $g$ can be used to herald the successful creation of an entangled pair \cite{Gisin10}. Two auxiliary photons $a_{\rm{H}}^\dag a_{\rm{V}}^\dag$\ket{0} in the same spatial mode $a$ but with orthogonal polarizations are sent through a tunable beam splitter with reflection coefficient $R$ (in intensity). This leads to the creation of two independent entangled states 
\begin{eqnarray}
\nonumber
&&\left(\sqrt{R} \-\ c_{\rm H}^\dag + \sqrt{1-R} \-\ out_{\rm H}^\dag\right) \ket{0} \otimes \\
\nonumber
&&\left(\sqrt{R} \-\ c_{\rm V}^\dag + \sqrt{1-R} \-\ out_{\rm V}^\dag\right)\ket{0}.
\end{eqnarray}
A Bell measurement is then performed on the modes $in$ and $c$ by detections in modes $d_{\pm} \propto c_{\rm H}+c_{\rm V} \pm in_{\rm H} \mp in_{\rm V}$ and $\tilde d_{\pm} \propto \pm c_{\rm H} \mp c_{\rm V} + in_{\rm H} + in_{\rm V}$ using the setup shown in Fig. \ref{fig QA}. In the ideal case, a twofold coincidence detection between $d_+$ and $\tilde d_-$ projects the modes $g$ and $out$ onto the maximally entangled state
\begin{equation}
\label{max_ent}
\frac{1}{\sqrt{2}}(g_{\rm{H}}^{\dagger} out_{\rm{H}}^{\dagger} + g_{\rm{V}}^{\dagger} out_{\rm{V}}^{\dagger}) \ket{0}.
\end{equation}
Note that the probability for the successful preparation of this state is $\frac{1}{4}pR(1-R).$ Since the twofold coincidences $d_+-\tilde d_+,$ $d_--\tilde d_+,$ $d_--\tilde d_-$ combined with the appropriate one-qubit transformation also collapse the modes $g$ and $out$ into the state (\ref{max_ent}), the overall success probability for the entangled-pair preparation is given by 
\begin{equation}
\label{proba_ampli}
P_{ampli}=pR(1-R).
\end{equation}
The described device thus serves as an heralding process for the successful creation of an entangled pair. If the modes $out$ and $g$ are further stored in quantum memories, this provides an on-demand entangled-pair source. \\

Note that the Bell measurement is different from the one originally presented in Ref. \cite{Gisin10}. Following the approach proposed e.g. in Ref. \cite{Chen07}, the Bell measurement here consists of a polarization beam splitter (PBS) in the $\pm45^o$ basis followed by a PBS in the H/V basis at each output (see \fig{\ref{fig QA}}). This setup automatically eliminates the vacuum in the memories. Indeed, in the ideal case where the single-photon sources and the detectors have unit-efficiencies, the only situation yielding vacuum in both memories is when no pair is emitted and both photons coming from the single photon sources are reflected at the BS. Since $c_{\rm H} c_{\rm V}$ leads to $\left(d_++d_-\right)^2-\left(\tilde d_+-\tilde d_-\right)^2,$ this event cannot produce the desired $d-\tilde{d}$ coincidence detection \cite{footnote_bunch}. \\

So far, we neglected the multi-pair emission from the pair source and we assumed unit-efficiency single-photon detectors and sources. Let us now relax these assumptions and see what are the effects of these imperfections on the photon-pair source. Consider first that the entangled-pair source produces two pairs with a probability proportional to $p^2$, as a spontaneous parametric down conversion source does, i.e.
\begin{eqnarray}
\rho_{pair} &=& (1-p-3/4p^2)\ket{0}\bra{0} + \\
\nonumber
&& \frac{p}{2} (g^\dagger_{\rm{H}} in^\dagger_{\rm{H}} + g^\dagger_{\rm{V}} in^\dagger_{\rm{V}})\ket{0}\bra{0}(g_{\rm{H}} in_{\rm{H}} + g_{\rm{V}} in_{\rm{V}})+\\
\nonumber
&&
\frac{p^2}{16} (g_{\rm{H}}^{\dagger} in_{\rm{H}}^{\dagger} + g_{\rm{V}}^{\dagger} in_{\rm{V}}^{\dagger})^2 \ket{0}\bra{0} (g_{\rm{H}} in_{\rm{H}} + g_{\rm{V}} in_{\rm{V}})^2.
\end{eqnarray}
Further consider that the single-photon sources have non-unit efficiencies, i.e produce a single photon with the probability $q$ only. The corresponding state is 
\begin{equation}
\rho_{single} = (1-q)\ket{0}\bra{0} + q a_{\rm{j}}^{\dagger} \ket{0}\bra{0}a_{\rm{j}},  
\end{equation}
where $j=\{{\rm H,V} \}.$ Moreover, consider that the single-photon detectors have a non-unit efficiency $\eta_D,$ i.e. the probability to get k clicks when a bunch of n photons impinge on a detector is given by $C_n^k \eta_D^k (1-\eta_D)^{n-k}$ where $C_n^k$ is the binomial coefficient. Let $\rho_0^{s}$ the density matrix describing the modes $g$ and $out$ once a successful Bell measurement is obtained. $\rho_0^{s}$ now contains components with multiple excitations. However, one can get rid of some of them under reasonable assumptions, namely 
\begin{subequations}
	\label{eqs approx}
	\begin{align}
		1 & > R \gg p,  \label{eq approx_Rp} \\
		1 & \gg 1-q. \label{eq approx_q}
	\end{align}	
\end{subequations}
These assumptions correspond to a photon pair source with a low double pair production and to single-photon sources with high efficiencies. In this regime, the (non-normalized) density matrix $\rho_0^{s}$ is well approximated by 
\begin{eqnarray}
		\rho_0^{s} & \approx & \alpha_0^s \; \ket{\phi^{g,out}_+}\bra{\phi^{g,out}_+} + \beta_0^s \; \ket{g_+}\bra{g_+} + \nonumber \\
							 && \gamma_0^s \; \ket{g_{2+}, out_{2\cdot} }\bra{g_{2+},out_{2\cdot}}  + \nonumber \\
							 && \delta_0^s \; \ket{g_{2+}, out_+ }\bra{g_{2+}, out_+},
	\label{eq rho0_s}	
\end{eqnarray}
where
\begin{subequations}
	\begin{align}
		\ket{\phi^{g,out}_+} &=\frac{1}{\sqrt{2}}(g_{\rm{H}}^{\dagger} out_{\rm{H}}^{\dagger} + g_{\rm{V}}^{\dagger} out_{\rm{V}}^{\dagger}) \ket{0} \\		
		\ket{g_+}\bra{g_+} &= \ket{ g_{\rm{H}} } \bra{ g_{\rm{H}} }+ \ket{ g_{\rm{V}} }\bra{ g_{\rm{V}} }	\\
		\ket{g_{2+}} &= ( (g^\dagger_{\rm{H}})^2 - (g^\dagger_{\rm{V}})^2 )\ket{0} \\
		\ket{out_{2\cdot}} &= \ket{out_{\rm{H}}  out_{\rm{V}}}	\\
		\ket{out_+}\bra{out_+} &= \ket{ out_{\rm{H}} } \bra{ out_{\rm{H}} }+ \ket{ out_{\rm{V}} }\bra{ out_{\rm{V}} }
	\end{align}	
\end{subequations}
and where the coefficients depend on the parameters $\eta_D, R, p, q$, ($T=1-R$) as follows
\begin{subequations}
	\begin{align}
		\alpha_0^s &= \frac{1}{4}\eta_D^2 pRTq^2	\\
		\beta_0^s &= \frac{1}{8} \eta_D^2 pRq  \left( 1-q + (1-\eta_D)Rq \right)	\\
		\gamma_0^s &= \left( \frac{\eta_D}{8} \right)^2 (pTq)^2	\\
		\delta_0^s &= \left( \frac{\eta_D}{8} \right)^2 p^2 Tq(1-q).
	\end{align}	
\end{subequations}
Note that the success probability $P_0^s$ for the preparation of the mixed state (\ref{eq rho0_s}) is equal to four times the trace of $\rho_0^{s}.$ (As before, four twofold coincidences lead to the same state using the appropriate unitary transformations.) The average waiting time for the successful photon-pair preparation is thus given by $T_0^s=\frac{1}{\gamma_{rep}P_{0}^s}$ where $\gamma_{rep}$ is the basic repetition rate of sources.\\

Let us briefly comment on the different contributions to the density matrix. First, the coefficients $\gamma_0^s, \delta_0^s$ scale as $p^2$ and are thus negligible with respect to $\beta_0^s$ and $\alpha_0^s$ when the pair emission is small $(p \ll 1).$ The term $\ket{g_+}\bra{g_+}$ is caused by the inefficiency of detectors and sources. Note that the imperfections coming from the detectors can be reduced by increasing the transmission of the tunable beam splitter if one is willing to lower the success probability. Note also that the ones coming from the single-photon sources cannot be lowered in this way. However, we show in what follows that the  errors ($\ket{g_+}\bra{g_+}$) resulting from both imperfect detectors and sources can be purified in the elementary link if one performs an entanglement swapping operation based on a two-photon detection. This is a major difference with the source presented in Ref. \cite{Sangouard08}, where a tradeoff between high fidelities and high creation rates leads inevitably to errors in the elementary links.  

\section{Entanglement creation in the elementary link}
\label{subsec ent creation}

\begin{center}
	\begin{figure}[h!]
  	\includegraphics[width=9cm]{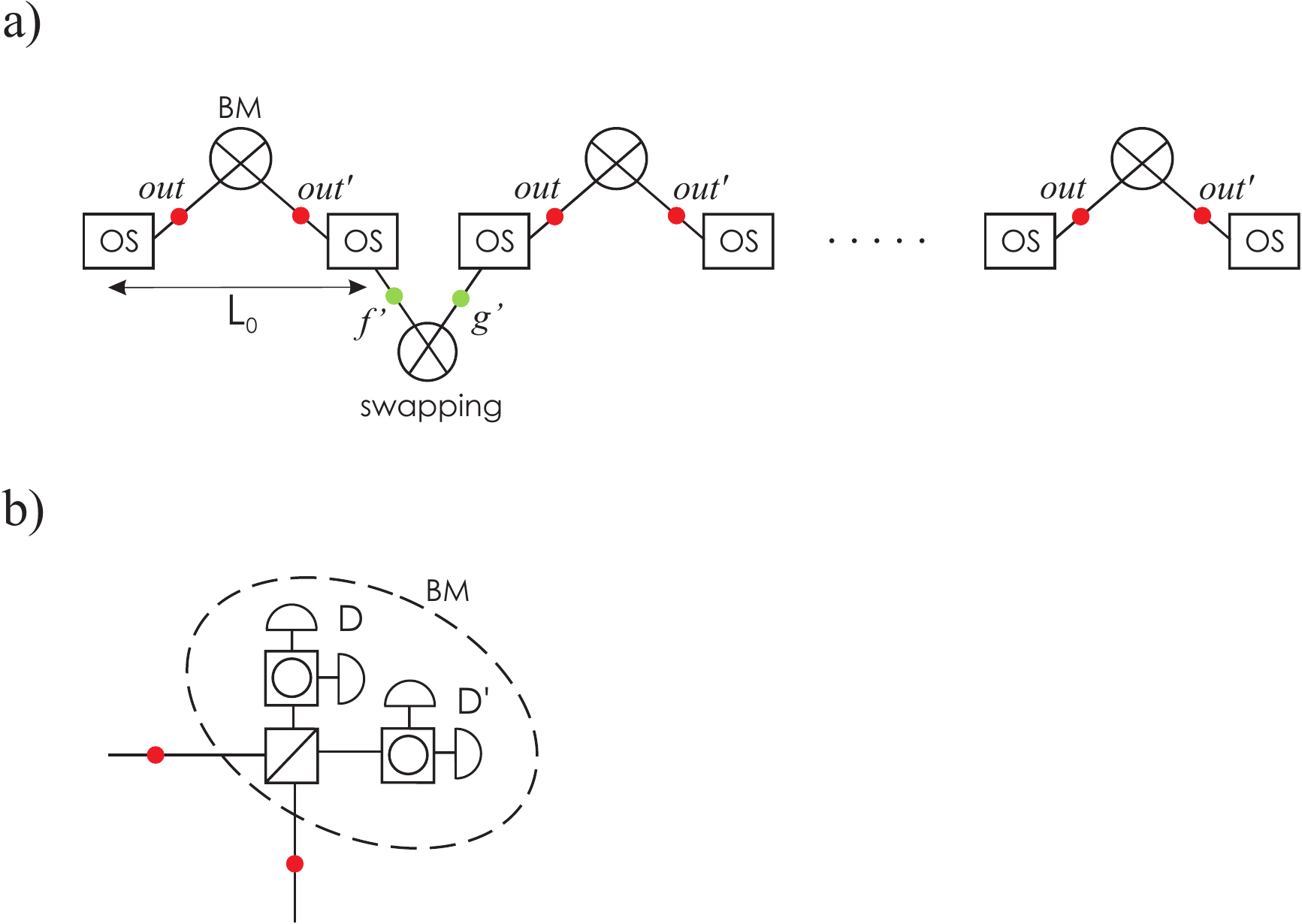} 
  	\caption{(Color online) a) A scheme of the whole repeater consisting of $2^n$ elementary links of length $L_0.$ Blocks labeled OS represent the {\it on site} blocks presented in \fig{\ref{fig QA}}. The entanglement in each elementary link is created by converting the $out$ atomic excitations  into optical modes and by performing subsequently  a Bell measurement (BM). The BM is represented here by the $\otimes$ symbol. When the entanglement is succesfully created in neighboring elementary links, they are then swapped using the same BM (swapping between modes $f'$ and $g'$ is shown; see the text for the details). b) Detail of the BM used in the elementary link and for the swappings. The BM consists of one PBS in the H/V basis followed by a PBS in $\pm 45^o$ at each output. } 
  	\label{fig Swapping}
 	\end{figure}
\end{center}

The entanglement in the elementary link is created by using one qubit amplifier at each location. The modes $out$ that are stored in atomic ensembles, are read out and the resulting photons are sent at a central station where a Bell measurement is performed (see \fig{\ref{fig Swapping}}). Let $out$ and $out'$ be the modes coming from the two remote locations. The Bell measurement consists in a projective measurement into the modes $D_\pm \propto out_{\rm H} \pm out'_{\rm V}$ and $D'_\pm \propto out'_{\rm H} \pm out_{\rm V}.$ It is realized using photon counting preceded by a PBS in the H/V basis followed by a $\pm 45^o$ PBS in each output of the H/V PBS. A twofold coincidence $D$-$D'$ projects the stored modes  $g$-$g'$ into 
\begin{eqnarray}
\rho_0^{g,g'} &=& \alpha_0 \; \ket{\phi^{g,g'}_+}\bra{\phi^{g,g'}_+} + \nonumber \\
&&\beta_0 \; (\ket{g_+, g'_{2+}}\bra{g_+, g'_{2+}} + \ket{g'_+, g_{2+}}\bra{g'_+, g_{2+}} ) + \nonumber \\
&& O(\ket{g_{2+}}\bra{g_{2+}},\ket{g'_{2+}}\bra{g'_{2+}}).
\label{eq rho0}
\end{eqnarray}
The coefficients $\alpha_0$ and $\beta_0$ are given in the Appendix (Eq. \ref{eqs coeffs_0}). What matters is that, first, the terms containing two excitations at each side (included in the $O(..)$ term) are negligible under the approximation (\ref{eqs approx}). Secondly, $\alpha_0 \propto  (\alpha_0^s)^2$ while $\beta_0$ is the sum of three contributions, the first one being proportional to $\alpha_0^s\delta_0^s,$ the second one $\propto \beta_0^s \gamma_0^s$ and the last one $\propto \alpha_0^s \gamma_0^s.$ Therefore, if the success probability for the pair emission is weak ($p \ll 1$), $\gamma_0^s$ and $\delta_0^s$ are negligible with respect to $\beta_0^s$ and $\alpha_0^s,$ meaning that the fidelity $\langle \phi^{g,g'}_+ |\rho_0| \phi^{g,g'}_+\rangle$ is very high. In other words, in the regime where the multi-pair emission can be neglected, the state created in the elementary link is maximally entangled, independently of the reflection coefficient $R.$ We now show that the quality of the entanglement is preserved after the entanglement swapping operations. 

\section{Entanglement swapping operations}
\label{subsec Swapping}

Consider two neighboring links, each of them described by the state (\ref{eq rho0}). Let $f,\; f'$ ($g,\; g'$) be the modes stored in the memories of the first (second) elementary link - see \fig{\ref{fig Swapping}}. The initial state is thus $\rho_0^{g,g'}\otimes\rho_0^{f,f'}.$ Consider that the prime modes $f',\; g'$ are located at the same location. The entanglement swapping operation consists in applying the Bell measurement used at the central station of each link (see Fig. \ref{fig Swapping} b)) once the atomic excitations corresponding to $f'$ and $g'$ are mapped into photonic modes. The Bell measurement succeeds when one gets a twofold coincidence between, for example $f'_{H}+g'_{V}$ and $g'_{H}+f'_{V}$. The resulting density matrix is of the form
\begin{eqnarray}
	\rho_1^{f,g} &=& \alpha_1 \; \ket{\phi^{f,g}_+}\bra{\phi^{f,g}_+} + \nonumber \\
	 && \beta_1 \; (\ket{f_+, g_{2+}}\bra{f_+, g_{2+}} + \ket{g_+, f_{2+}}\bra{g_+, f_{2+}} ) + \nonumber \\
	 && \gamma_1 \; \ket{f_+, g_+}\bra{f_+, g_+} + \nonumber \\
	 && O(\ket{g_{2+}}\bra{g_{2+}},\ket{f_{2+}}\bra{f_{2+}}),
	\label{eq rho1}
\end{eqnarray}
where $\alpha_1$, $\beta_1$ and $\gamma_1$ are coefficients given in the Appendix (see Eq. (\ref{eqs coeffs_1})). As before, the $\beta_1,$ $\gamma_1$ and $O(..),$ terms can be neglected in the regime where the inequalities (\ref{eqs approx}) are satisfied, and in this case, the distributed state reduces to a maximally entangled state. \\

The entanglement swapping operations are repeatedly performed until the entanglement is extended to the desired distance. One can show that the state $\rho_k$ obtained after the k$^{\rm th}$ entanglement swapping 
has the same form that $\rho_1^{f,g}.$ The corresponding coefficients $\alpha_k,$ $\beta_k$ and $\gamma_k$
are explicitly given in the appendix (see Eq. (\ref{eqs coeffs_k})). Therefore, the final state does not contain neither vacuum nor components with only one excitation in the regime described by the inequalities  (\ref{eqs approx}). Our protocol is the only one known to us based on atomic ensembles and linear optics which does not require any final post-selection. This provides not only unequaled performances but also unique opportunities e.g. for device independent quantum key distribution over long distances.

\section{Performance}
\label{subsec Results}

The explicit knowledge of the density matrix at each step allows us to calculate the success probabilities for the entanglement creation and for the entanglement swapping operations, and thus to access the performance of the proposed protocol using the formula (\ref{eq Ttot_basic}). Note that this formula is strictly valid if the source preparation time is negligible compared to the communication tim, i.e. $T_{0}^s \ll \frac{L_0}{c}.$ Otherwise, $\frac{L_0}{c}$ has to be replaced by $T_{0}^s + \frac{L_0}{c}.$\\

For concreteness, consider the distribution of entanglement over $L$=1000 km using memories and detectors with efficiencies $\eta_M = \eta_D = 0.9.$ For a given quality of single-photon sources (fixed $q$), we look for the values of $p$ and $R$ for each $n$ leading to a fidelity of the distributed state at least equal to 0.9. We then keep the triplet $p,\; R,\;n$ that minimize the average time for the distribution of one entangled pair. Note that we restrict the maximum number of nesting level to be $n\leq 4$ (16 links) in such a way that one can reasonably make the assumption that purification is not necessary to correct the imperfections coming potentially from swapping operations \cite{Sangouard11}. \\

\begin{table}[h!]
	\centering
	\begin{tabular}{c c c c c c}
	
	 	$T_{tot}$ [s] & $p$ & $q$ & $R$ & $F$ & $\gamma_{rep}$ [MHz] \\ \hline	 		
	 	7.4 & $6 \cdot 10^{-4}$ & 1 & 0.12 & 0.96 & 60  \\
  	7.8 & $3.6 \cdot 10^{-3}$ & 1 & 0.23 & 0.9 & 6  \\ 
	19.2 & $6 \cdot 10^{-4}$ & 0.66 & 0.17 & 0.96 & 60 \\ \hline   
	\end{tabular}
  \caption{Performance of the proposed quantum repeater based on qubit amplifiers. The results correspond to the distribution of entangled pairs over $L=1000$ km, with memory and detector efficiencies $\eta_M=\eta_D=0.9$ using 16 links ($n=4$). $T_{tot}$ is the average time for the distribution of one entangled pair, $p$ ($q$) is the probability with which the pair source (single-photon sources) used within the qubit amplifier emits a photon-pair (emit a single photon). $R$ is the beam splitter reflectivity. $F$ is the fidelity of the distributed state and $\gamma_{rep}$ is the basic repetition rate of sources.}
  \label{tab Results}
\end{table}

The results are shown in the first row of the Table \ref{tab Results} \cite{note}. With ideal single-photon sources ($q=1$), the average time to distribute one entangled pair with the proposed scheme ($T_{tot} \approx 7.4$ s) is very close to what could be obtained ideally, with hypothetical sources producing maximally entangled photon pairs on-demand ($T_{tot} \approx 6$ s, c.f. Section \ref{sec QR local} ). Let us mention that a source of single-photons with near-unity collection efficiency has been reported very recently \cite{Lee11}. \\

Note that for these values, equality between the source preparation time $T_{0}^s $ and the communication time $L_0/c$ is reached for a basic repetition rate of $\gamma_{rep} = 60$ MHz. Current spontaneous-Raman-emission based sources with atomic gases are well suited for demonstrating the proposed source. However, current repetition rates $\gamma_{rep}$ are of the order of a few MHz \cite{Chou07}. The second row of the Table \ref{tab Results} shows that if one chooses a weaker transmission for the tunable beam splitter and a higher pair production, one can distribute good quality entanglement with high entanglement distribution rates while significantly reducing the required basic repetition rate. (In this case, equality between the source preparation time $T_{0}^s $ and the communication time $L_0/c$ is reached for a basic repetition rate of $\gamma_{rep} = 6$ MHz, a level likely achievable with present-day experiments.)\\

It is interesting to compare the performance of the proposed protocol with the one presented in Ref. \cite{Sangouard08}. The latter is the most efficient quantum repeater based on atomic ensembles and linear optics known to us. It takes $19$ s in average to distribute one entangled pair with resources comparable to the ones presented in the first row of the Table \ref{tab Results}. The third row of this Table shows that the protocol based on heralded qubit amplifiers achieves an advantage over the one of Ref. \cite{Sangouard08} as soon as $q \geq 0.7.$ Note, in the one hand, that if the single-photon sources are performed by combining a pair source and a quantum memory \cite{Sangouard11}, the proposed protocol requires the storage of six modes for the implementation of one on-demand entanglement source while the one of Ref. \cite{Sangouard08} needs four only. On the other hand, the proposal of Ref. \cite{Sangouard08} achieves high fidelities only with postselections, even taking memory and detector imperfections into account. We emphasize that the here presented quantum repeater allows one the distribution of highly entangled states without post-selection. \\

Before we conclude, we would like to add three comments. First, an additional speed up could be achieved, at least in principle, using a temporal multiplexing \cite{Simon07}. However, regarding the required repetition rates $\gamma_{rep},$ a useful multiplexing would require quantum memory with prohibitively large bandwith and multimode capabilities. Secondly, the modes $out$ have been considered to be at telecommunication wavelength where the attenuation of optical fibers is low. Therefore, our proposal requires either memories suited for telecom wavelengths or an efficient frequency conversion from the storable wavelength to the telecom wavelength. The former has recently been realized in a rare-earth-doped material \cite{Lauritzen10}, though not with the desired efficiency. Several proof-of-principle experiments have also been demonstrated recently for the latter, either using parametric down-convertion \cite{down_conv} (albeit not with quantum memory compatible photons) or four wave mixing in cold atomic ensembles \cite{Radnaev10}. Finally, significant speed-ups could be obtained if the swapping operations were performed deterministically. This might be done using either quantum memories based on single trapped ions \cite{Sangouard09} or Rydberg transitions \cite{Rydberg}, or entanglement based on continuous variables \cite{Sangouard10}.\\

\section{Conclusion}
\label{sec Conclusion}

We started this paper by reminding that the local generation of high-fidelity entangled pairs and the heralded creation of entanglement in each basic link are required for quantum repeaters to achieve high entanglement distribution rates. We then proposed a source, inspired by the heralded qubit amplifier, to produce on-demand entangled pairs. We showed how the potential errors resulting from the use of non-unit efficiency devices, are purified by the entanglement swapping operation serving as heralding for the preparation of entanglement in the elementary link. The quality of the entanglement is preserved even after many entanglement swapping operations such that the proposed quantum repeater does not require post-selection as usually done in repeaters with atomic ensembles and linear optics. This leads to entanglement distribution rates similar to the ones that would be obtained from ideal photon-pair sources. Furthermore, this opens a way for the realization of device-independent quantum key distribution over very long distances.\\
~\\
We thank M. Afzelius and N. Gisin for interesting discussions. We gratefully acknowledge support by the EU projects Qscale and Qessence and from the Swiss NCCR QSIT.

\section*{APPENDIX}
\label{App}

\subsection*{Coefficients calculation}

 \renewcommand{\theequation}{A-\arabic{equation}}
  \setcounter{equation}{0}  

Lets denote the success probability at each site as $P_0^s$, which is defined as

\beq
	P_0^s = 4 {\rm Tr} \left( \rho_0^s \right) = 4 \left(\alpha_0^s + 2\beta_0^s + 4\gamma_0^s + 8\delta_0^s\right),
\eeq
where $\rho_0^s$ is given by \eq{\ref{eq rho0_s}}. The coefficients of the \emph{non normalized} density matrix $\rho_0$ (\eq{\ref{eq rho0}}) after the Bell measurement in the elementary link read

\begin{subequations}
	\label{eqs coeffs_0}
	\begin{align}
		\alpha_0 &= \frac{1}{({\rm Tr} \left( \rho_0^s \right))^2}\frac{\eta_{DMt}^2}{8} (\alpha_0^s)^2 \\
		\beta_0 &= \frac{1}{({\rm Tr} \left( \rho_0^s \right))^2}\frac{\eta_{DMt}^2}{2} \left[ \frac{\frac{\alpha_0^s}{2} \delta_0^s + \beta_0^s \gamma_0^s}{2} + \alpha_0^s \gamma_0^s (1-\eta_{DMt}) \right],
	\end{align}	
\end{subequations}
where $\eta_{DMt} = \eta_D \eta_M \eta_t$ is the product of the detector, memory and channel transmission efficiencies. We can now continue in an analogous fashion for all density matrices. The success probability of entanglement creation in the elementary link is
\beq
 P_0 = 4 {\rm Tr} \left( \rho_0 \right) = 4 \left(\alpha_0 + 16\beta_0\right),
\eeq
we can then write for the coefficients of the non normalized density matrix $\rho_1^{f,g}$ (\eq{\ref{eq rho1}}) after the first swapping

\begin{subequations}
	\label{eqs coeffs_1}
	\begin{align}
		\alpha_1 &= \frac{1}{({\rm Tr} \left( \rho_0 \right))^2} \frac{\eta_{DM}^2}{8} \alpha_0^2 \\
		\beta_1 &= \frac{1}{({\rm Tr} \left( \rho_0 \right))^2} \eta_{DM}^2 (1-\eta_{DM}) \beta_0 \left[ \alpha_0 + 8 \beta_0 (1-\eta_{DM}) \right] \\
		\gamma_1 &= \frac{1}{({\rm Tr} \left( \rho_0 \right))^2} \frac{\eta_{DM}^2}{8} \beta_0 \left[ \alpha_0 + 16\beta_0 (1-\eta_{DM}) \right],
	\end{align}	
\end{subequations}
where $\eta_{DM} = \eta_D \eta_M$. As discussed in the text, under the assumptions \eq{\ref{eqs approx}}, each following swapping will yield the density matrix of the form \eq{\ref{eq rho1}}. We can thus write recurrent relations for the non normalized coefficients of the density matrix $\rho_k$ obtained after the $k-$th swapping ($k>1$)

\begin{subequations}
	\label{eqs coeffs_k}
	\begin{align}
		\alpha_k =& \frac{1}{({\rm Tr} \left( \rho_{k-1} \right))^2} \frac{\eta_{DM}^2}{8} \alpha_{k-1}^2 \\
		\beta_k =& \frac{1}{({\rm Tr} \left( \rho_{k-1} \right))^2} \frac{\eta_{DM}^2}{4} [ \beta_{k-1} (\alpha_{k-1} + 2\beta_{k-1}) + \nonumber \\
		& 4\gamma_{k-1}(1-\eta_{DM})(\alpha_{k-1} + 4\beta_{k-1}) + 2(4\gamma_{k-1}(1-\eta_{DM}))^2 ] \\												
		\gamma_k =& \frac{1}{({\rm Tr} \left( \rho_{k-1} \right))^2} \frac{\eta_{DM}^2}{8} \gamma_{k-1} \left[ \alpha_{k-1} + 4\beta_{k-1} + 16\gamma_{k-1} (1-\eta_{DM}) \right],
	\end{align}	
\end{subequations}
where
\beq
 P_{k-1} = 4{\rm Tr} \left( \rho_{k-1} \right) = 4\left(\alpha_{k-1} + 4\beta_{k-1} + 16\gamma_{k-1}\right)
\eeq
is the success probability of the $k-$th swapping.


\begin{thebibliography}{3}
\bibitem{note_fid} Note that in the case where the source emits a maximally entangled state with unit efficiency, the final post-selected state is maximally entangled.
\bibitem{footnote_bunch} Note that in Ref. \cite{Pitkanen11}, the authors have presented an interesting alternative which benefits from the bunching to avoid the vacuum component in the heralded state.
\bibitem{note} An obvious remark is that one has to verify whether the used approximations \eq{\ref{eqs approx}} still hold. If they are not fully respected, one has to verify the complete density matrix at each site $\rho_0^s$ and in the elementary link $\rho_0$. We verified that for the results shown in Table \ref{tab Results} the weight of the neglected components is at least $10^3$ times smaller than the weight of the entangled state (which is the leading term) and we thus consider the approximation valid.
\end{thebibliography}
\end{document}